\documentclass[acmtog]{acmart}
\AtBeginDocument{%
  }

\usepackage{soul}
\sethlcolor{yellow}
\definecolor{revcolor}{RGB}{0,0,200} % 你也可以改成别的颜色

\usepackage{multirow}
\usepackage{dsfont}
\usepackage{etoolbox}
\usepackage{color}

\makeatletter
\DeclareRobustCommand*\cal{\@fontswitch\relax\mathcal}
\makeatother

\newif\ifshowedits

\newcommand{\addeditor}[3]{%
  \definecolor{#1color}{rgb}{#3}
  \expandafter\newcommand\csname #1\endcsname[1]{%
  \ifshowedits
    {\color{#1color} ##1}%
  \else
    {##1}%
  \fi
  }%
  \expandafter\newcommand\csname #1rmk\endcsname[1]{%
  \ifshowedits
    {\color{#1color} {\bf [#2: ##1]}}
  \fi
  }%
  \expandafter\newcommand\csname #1rpl\endcsname[2]{%
  \ifshowedits
    {\color{#1color} ##1 \sout{##2}}
  \else
    {##1}
  \fi
  }%
}

\newcommand{\createtextvar}[1]{
  \expandafter\newcommand\csname #1\endcsname{%
  {\text{#1}}
}%
}
\newcommand{\mycomment}[1]{}

\newcommand{\vcomment}[1]{}

\usepackage{colortbl}

\definecolor{yellow}{rgb}{1, 1, 0.7}
\definecolor{orange}{rgb}{1, 0.85, 0.7}
\definecolor{tablered}{rgb}{1, 0.7, 0.7}
\definecolor{red}{rgb}{1, 0, 0}

\definecolor{tablethree}{rgb}{0.7, 1, 1}
\definecolor{tabletwo}{rgb}{0.7, 0.85, 1}
\definecolor{tableone}{rgb}{0.7, 0.7, 1}

\newcommand{\best}{\cellcolor{tablered}}
\newcommand{\sbest}{\cellcolor{orange}}
\newcommand{\tbest}{\cellcolor{yellow}}

\newcommand{\bestbis}{\cellcolor{tableone}}
\newcommand{\sbestbis}{\cellcolor{tabletwo}}
\newcommand{\tbestbis}{\cellcolor{tablethree}}

\usepackage{tikz}
\usepackage{tikz-3dplot}
\usetikzlibrary{calc,arrows.meta, intersections}
\usetikzlibrary{positioning, shapes.misc}
\usepackage{tkz-euclide}
\usetikzlibrary{patterns, shapes.geometric, quotes, backgrounds}

\definecolor{cmarkgreen}{RGB}{46, 139, 87}
\definecolor{xmarkred}{RGB}{178, 34, 34}

\usepackage{fontspec} % 必须使用 XeLaTeX 编译，用于调用系统字体
\usepackage{pifont} % 提供 \ding{51}/\ding{55}，避免与 acmart/newtxmath 冲突
\usepackage{xcolor}
\definecolor{darkred}{RGB}{130, 0, 0} % 稍微调深一点，更有质感
\AtBeginDocument{%
  }

\setcopyright{acmlicensed}
\copyrightyear{2026}
\acmYear{2026}
\acmDOI{XXXXXXX.XXXXXXX}
\acmISBN{978-1-4503-XXXX-X/2018/06}
\usepackage{array}
\usepackage{enumitem}
\acmSubmissionID{898}

\acmJournal{TOG}

\begin{document}

\title{Distance Field Rasterization for End-to-End Mesh Reconstruction}


%% Authors
\author{Jinkai Cui}
\email{cuijk@mail.ustc.edu.cn}
\orcid{0009-0007-3115-2807}
\affiliation{%
  \institution{University of Science and Technology of China}
  \country{China}
}

\author{Kaiwen Song}
\email{SA21001046@mail.ustc.edu.cn}
\orcid{0009-0007-1199-5380}
\affiliation{%
  \institution{University of Science and Technology of China}
  \country{China}
}

\author{Chumeng Niu}
\email{niuchumeng@mail.ustc.edu.cn}
\orcid{0009-0004-0098-6137}
\affiliation{%
  \institution{University of Science and Technology of China}
  \country{China}
}

\author{Juyong Zhang}
\email{juyong@ustc.edu.cn}
\authornote{Corresponding author (\href{mailto:juyong@ustc.edu.cn}{juyong@ustc.edu.cn}).}
\orcid{0000-0002-1805-1426}
\affiliation{%
  \institution{University of Science and Technology of China}
  \country{China}
}

\renewcommand{\shortauthors}{Cui et al.}

\begin{abstract}
Rasterization based methods have recently enabled high-quality novel view synthesis at real-time rates, but their underlying volumetric primitives do not expose a direct, globally consistent surface representation, leaving surface extraction to heuristic post-processing. In contrast, implicit signed distance field (SDF) methods provide well-defined surfaces but are typically optimized with computationally expensive ray marching. We propose \textbf{SDFRaster}, a rasterizable SDF representation that bridges this gap by combining the efficiency of rasterization with signed distance field for end-to-end mesh reconstruction. Starting from a Delaunay tetrahedralization, we optimize a continuous SDF over a tetrahedral grid and render it efficiently by rasterizing tetrahedra and alpha-compositing their contributions. We further integrate differentiable Marching Tetrahedra into the optimization loop, enabling end-to-end mesh reconstruction without post-processing mesh extraction. Experiments on DTU and Tanks and Temples demonstrate that SDFRaster achieves higher-quality and more complete surface reconstructions with lower storage cost than state-of-the-art approaches. Project page: \href{https://ustc3dv.github.io/SDFRaster/}{ustc3dv/SDFRaster}
\end{abstract}



\begin{teaserfigure}
  \includegraphics[width=\textwidth]{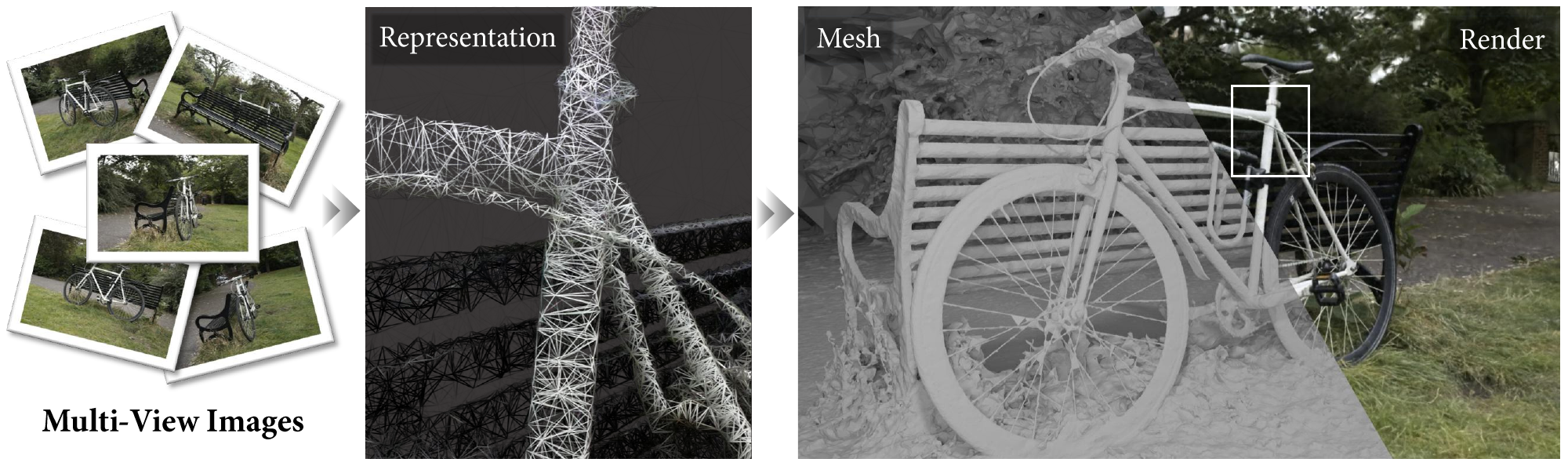}
  \caption{Our method introduces a rasterizable SDF representation for end-to-end mesh reconstruction from multi-view images. With this representation, our method achieves high-quality mesh reconstruction (right). The middle column shows the optimized tetrahedral mesh which supports the SDF. }
  \Description{ }
  \label{fig:teaser}
\end{teaserfigure}


\maketitle

\section{Introduction}
\label{sec:intro}

Mesh reconstruction is a fundamental problem in computer graphics and computer vision, traditionally addressed by multi-view stereo (MVS) methods~\cite{colmap, yao2018mvsnet,schoenberger2016mvs,yu_2020_fastmvsnet,kazhdan2013screened} that recover explicit geometry from calibrated images, yet often suffer from noise sensitivity and limited robustness in complex scenes. Recently, neural implicit representations model surfaces~\cite{idr, neus, neus2, neuralangelo,volsdf,cai2022ndr,jiang2022selfrecon,Wang2021PriorGuidedM3} as continuous functions and improve completeness and smoothness, but they typically require dense sampling along rays and frequent network evaluations during training, making optimization computationally expensive. Building upon these advances, rasterization based methods~\cite{3dgs, svraster, radiancemeshes} have emerged as an efficient alternative, where fast GPU rasterization enables rapid rendering and optimization. Leveraging this computational advantage, many recent methods~\cite{2dgs, pgsr, svrecon} reconstruct surfaces efficiently, making rasterization based representations a compelling choice when reconstruction efficiency and scalability are primary concerns.

Although rendering quality and efficiency are high, mainstream rasterization based methods typically rely on volumetric primitives, which do not expose a well-defined surface representation such as signed distance fields. As a result, mesh reconstruction is often delegated to post-processing such as depth fusion, which can introduce noise, incompleteness, and accuracy degradation. On the other hand, the lack of a well-defined surface representation leads to weak inherent multi-view geometric consistency. Consequently, existing 3DGS-based surface reconstruction methods typically promote consistency only indirectly, for example by adding multi-view regularizers~\cite{pgsr,vcrgs}, introducing architectural inductive biases~\cite{2dgs, gaussiansurfel, pgsr}, or adopting dual-branch designs that co-optimize 3DGS with a surface branch via mutual supervision~\cite{milo, gsdf}. However, these constraints typically act on intermediate cues, such as depths and normals rendered by 3DGS, rather than directly supervising the underlying geometry, thereby weakening the coupling between the optimization objective and the final surface quality.

These limitations call for a representation that combines the well-defined surface geometry of SDF with the efficiency of rasterization. To this end, we propose \textbf{SDFRaster}, a rasterizable SDF representation for end‑to‑end mesh reconstruction. Building on the representation of Radiance Meshes~\cite{radiancemeshes}, we discretize the 3D scene using a Delaunay tetrahedralization and parameterize the SDF at the tetrahedral vertices. Within each tetrahedron, the SDF is obtained by linear interpolation of the vertex values, resulting in a continuous, piecewise-linear signed distance field over the scene. Thanks to this multi-view-consistent geometry, whose zero level set directly defines the surface, SDFRaster avoids the post-processing used in volumetric primitive methods and enables direct mesh extraction from the optimized geometric field. To render and optimize the signed distance field efficiently, we map the SDF to volumetric opacity and employ a rasterization-based differentiable renderer that rasterizes tetrahedra, computes intersections between rays and tetrahedra, and alpha-composites their contributions. This removes the need for per-ray dense sampling in ray marching renderers and enables fast, scalable optimization on modern GPU architectures.

Moreover, to better align field optimization with mesh quality, we extract meshes during optimization and couple two complementary geometry representations. Specifically, we integrate differentiable Marching Tetrahedra~\cite{marchingtetrahedra} into the optimization process, enabling surface-based regularization and geometry consistency losses directly on the extracted mesh. In addition, we render depth and normals from the SDF field and enforce their consistency with those rendered from the extracted mesh, tightening the link between appearance fitting and surface fidelity. Finally, we propose a surface-centric adaptive resolution strategy that concentrates representation capacity near the surface, improving local geometric detail while keeping the extracted meshes compact. Extensive experiments show that SDFRaster is over 6$\times$ faster than implicit SDF baselines, achieves the lowest Chamfer distance among explicit baselines on DTU, and maintains competitive F1 on TnT while producing compact meshes (about 3$\times$ smaller in storage than meshes extracted via TSDF fusion). 

In summary, the main contributions of this paper include:
\begin{itemize}
  \item We propose a Delaunay tetrahedral SDF representation that supports efficient rasterization based differentiable rendering and provides well-defined geometry.
  \item We integrate differentiable Marching Tetrahedra into optimization to extract meshes on the fly, enabling end‑to‑end mesh reconstruction and avoiding reliance on post‑processing.
  \item We adapt representation capacity near the evolving zero-level set to recover fine details while keeping the extracted meshes compact and scalable to large scenes.
\end{itemize} 
\section{Related Work}
\label{sec:related}

\subsection{Novel View Synthesis}

Neural radiance fields~\cite{nerf} model scenes as continuous volumetric functions optimized with differentiable volume rendering. Subsequent work improves quality and efficiency through compact parameterizations such as feature grids and tensor factorization, and through antialiasing or unbounded scene modeling~\cite{instantngp,plenoxels,Reiser2024SIGGRAPH,tensorf,dvgo,mipnerf360,zipnerf,fastnerf,Song2024City,Gao2022nerfblendshape}. These methods achieve strong novel view synthesis, but often require extensive training time.

Explicit rendering pipelines have shifted NVS toward real-time synthesis by rasterizing primitives. 3DGS~\cite{3dgs} represents scenes with explicit Gaussian primitives and achieves high-quality views at interactive rates, and variants~\cite{scaffoldgs,stoptethepop,2dgs} improve efficiency and stability while preserving the splatting pipeline. Recently, explicit representations not based on Gaussian splats have also appeared, such as sparse voxel rasterization~\cite{svraster} and cell-complex renderers~\cite{radiantfoam,kulhanek2023tetranerf,radiancemeshes,structurefield} that organize space into voxels or polyhedral cells for rasterization or ray tracing. These methods offer efficient NVS, but their primary goal remains view synthesis rather than accurate surface reconstruction.

\subsection{Neural Surface Reconstruction}

Classical multi-view reconstruction pipelines estimate depth or point clouds and then fuse them into a surface using volumetric fusion~\cite{yao2018mvsnet,schoenberger2016mvs,yu_2020_fastmvsnet} or implicit fitting~\cite{poissonrecon,kazhdan2013screened}. While effective in controlled settings, these pipelines are sensitive to noisy depth and view-dependent artifacts, and the final mesh quality is tied to the fusion resolution and coverage.

Neural surface reconstruction instead learns implicit fields and extracts their zero level set, while some methods based on discrete structures explore more direct mesh extraction and optimization~\cite{dmtet,chen2022ndc,flexicubes,Binninger:TetWeave:2025}. Early neural implicit methods such as IDR~\cite{idr} build on differentiable rendering to model surfaces implicitly. VolSDF~\cite{volsdf} and NeuS~\cite{neus} convert SDF to opacity and render via per‑ray volumetric integration with dense point sampling. This dense sampling makes training costly and leaves efficient large-scale reconstruction challenging~\cite{neuralangelo,2dgs}. Other variants improve robustness or reconstruction quality with stronger regularization, geometric priors, or training strategies~\cite{neus2,geoneus,monosdf,yu2022sdfstudio}. Neuralangelo leverages multi-resolution hash encodings with coarse-to-fine training and achieves high-quality surfaces~\cite{neuralangelo}. Recent work reduces the cost of volume rendering by adopting hybrid structures such as regular grids and factorized grids~\cite{voxurf,plenoxels,dvgo,tensorf}. These grid-based methods improve efficiency while preserving the implicit formulation, but they still rely on ray marching and remain expensive.

\subsection{Primitive-based Surface Reconstruction}

Gaussian splatting enables fast rendering and has motivated surface reconstruction on top of 3DGS~\cite{3dgs}. One line of work~\cite{2dgs,sugar,pgsr,dnsplatter} aligns splats with surfaces using planar or normal-aware regularization, then extracts meshes from rendered depth cues via TSDF fusion~\cite{kinectfusion,tsdf}. Another line strengthens geometric regularization within the splatting pipeline with depth/normal consistency, planar constraints, or higher-order primitives~\cite{li2024dngaussian,zhang2024rade,chen2024vcr,zhang2024quadratic,li2025gaussianudf}. In both cases, geometry is recovered after training and still relies on depth-fusion post-processing or overlapping-primitive cues, so surface quality is limited by multi-view depth consistency and depth accuracy, often yielding incomplete or overly dense geometry.

Hybrid methods jointly optimize Gaussian primitives with implicit distance fields. GSDF~\cite{gsdf}, GSurf~\cite{gsurf}, GS-Pull~\cite{zhang2024gspull}, and 3DGSR~\cite{3dgsr} introduce SDF supervision or gradient pulling toward the zero level set, and then extract meshes from the learned field, which improves geometric consistency but introduces a dual representation and additional optimization complexity.

More recent approaches reduce the gap between rendering and geometry with tighter extraction pipelines or mesh-based reconstruction~\cite{Held2025MeshSplatting}. GOF constructs Gaussian opacity fields and extracts meshes on a tetrahedral grid induced by Gaussians, enabling adaptive and compact meshes without TSDF fusion~\cite{gof}. MILo extracts mesh during training to reduce discrepancies between the rendered field and the extracted surface~\cite{milo}. These approaches couple surface extraction more directly to the optimized representation.
 
\section{Method}
\label{sec:method}
\begin{figure*}[t]
      \centering
      \includegraphics[width=\textwidth]{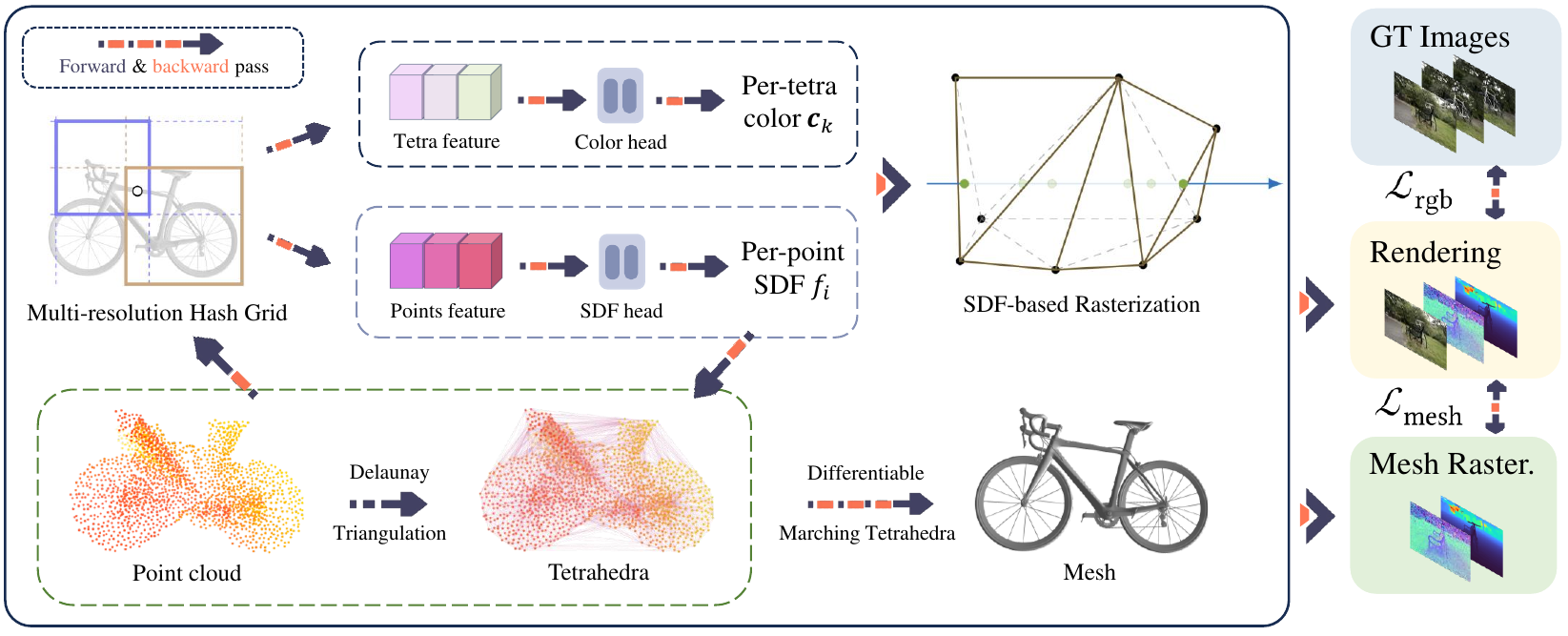}
      \caption{\textbf{Overview of SDFRaster}. We learn a continuous SDF on a Delaunay tetrahedral grid, using a shared multi-resolution hash encoder to predict SDF values at vertices and appearance per tetrahedron. We render the images by rasterizing tetrahedra and alpha-compositing SDF-derived opacities. We apply differentiable Marching Tetrahedra on the tetrahedral grid with the learned SDF values to extract meshes in the optimization, enabling end-to-end mesh reconstruction.}
      \label{fig:pipeline}
\end{figure*}

We propose \textbf{SDFRaster}, a rasterizable SDF framework that reconstructs meshes from calibrated multi-view images (see Fig.~\ref{fig:teaser}). Starting from a scene bounding volume, a Delaunay tetrahedralization is built as the geometric grid. A continuous SDF is learned over the tetrahedral complex and rendered by rasterizing tetrahedra with opacity-based alpha compositing. Meshes are extracted during training via differentiable Marching Tetrahedra, enabling geometry consistency losses without fixed‑resolution fusion. See Fig.~\ref{fig:pipeline} for a visual
overview of our method.
\subsection{Representation}
\label{sec:method_representation}

We maintain a set of vertices $\mathcal{V}=\{ \mathbf{v}_i \in \mathbb{R}^3 \}$ and compute a Delaunay tetrahedralization $\mathcal{T}$ over $\mathcal{V}$, forming a tetrahedral cell complex that partitions the scene into non-overlapping tetrahedra. This grid could provide an injective parameterization of the 3D scene, which is necessary to represent a signed distance function $f:\mathbb{R}^3\rightarrow\mathbb{R}$. To avoid discontinuities from directly optimizing discrete values, we parameterize the SDF with a compact network using multi-resolution hash encoding $E_\gamma$ ~\cite{instantngp,neuralangelo} queried on the tetrahedral grid. The SDF and appearance branches share the same $E_\gamma(\cdot)$ but use separate heads.

\textbf{Geometry.} We encode geometry as a signed distance field over the tetrahedral complex. Concretely, we predict a SDF value at each grid vertex and extend it to the interior of every tetrahedron via barycentric interpolation. The SDF value at vertex $\mathbf{v}_i$ is obtained by a dedicated SDF head $H_{\text{sdf}}$ that takes the hash-encoded feature and the raw vertex coordinates as input:
\begin{equation}
f_i \;=\; H_{\text{sdf}}\!\left(\left[E_{\gamma}(\mathbf{v}_i),\, \mathbf{v}_i\right]\right),
\label{eq:vertex_sdf_query}
\end{equation}
where $[\cdot,\cdot]$ denotes concatenation.
For any point $\mathbf{x}$ inside tetrahedron $t=(i_0,i_1,i_2,i_3)$ with barycentric coordinates $\lambda_k(\mathbf{x})$, where $k=0,1,2,3$,  satisfying $\sum_{k=0}^{3}\lambda_k(\mathbf{x})=1$ and $\lambda_k(\mathbf{x})\ge 0$, we define a continuous piecewise-linear SDF:
\begin{equation}
f(\mathbf{x}) \;=\; \sum_{k=0}^{3} \lambda_k(\mathbf{x})\, f_{i_k}.
\label{eq:pl_sdf}
\end{equation}
This yields a globally defined, piecewise-linear SDF over the tetrahedral complex, which we use for differentiable rendering (Sec.~\ref{sec:method_rendering}) and surface extraction. Notably, because the network is evaluated only at grid vertices and interior values are obtained via interpolation, our formulation substantially reduces the number of SDF network queries relative to methods that evaluate an MLP at every sample point, such as NeuS~\cite{neus}. Moreover, once the vertex SDF values are queried, the geometry can be stored and processed as a purely explicit representation.

\textbf{Appearance.} For each tetrahedron $t_k$, we choose the centroid as the representative query point $\mathbf{O}_k$. Attaching appearance directly to vertices is unstable under Delaunay topology changes; querying a representative point per tetrahedron yields stable per-cell attributes~\cite{radiancemeshes}. Using a shared multi-resolution hash encoder $E_{\gamma}(\cdot)$~\cite{instantngp} and an appearance head, we predict a view-dependent base color $\mathbf{c}^0_k(\mathbf{d})$ and a linear color gradient $\nabla \mathbf{c}_k$ at $\mathbf{O}_k$, and define the color at any interior point $\mathbf{p}\in t_k$ as a first-order model around $\mathbf{O}_k$:
\begin{equation}
  \mathbf{c}_k(\mathbf{p},\mathbf{d}) \;=\; \mathbf{c}^0_k(\mathbf{d}) \;+\; \nabla \mathbf{c}_k \cdot 
    \bigl(\mathbf{p}-\mathbf{O}_k\bigr)\,.
  \label{eq:linear-colour}
\end{equation}
Here $\mathbf{c}^0_k(\mathbf{d})$ is view-dependent and $\nabla \mathbf{c}_k$ is a per-tetrahedron linear color gradient predicted by the appearance head. Given the SDF representation anchored on the tetrahedral complex, we next describe a differentiable rendering procedure that rasterizes the piecewise-linear signed distance field.

\subsection{Rendering}
\label{sec:method_rendering}

Following Radiance Meshes~\cite{radiancemeshes}, we adopt a rasterizable volume rendering framework over the tetrahedral grid. The signed distance field is converted to opacity using the SDF-to-opacity mapping function in NeuS~\cite{neus}. For each pixel, we consider the corresponding camera ray $\mathbf{r}(t)=\mathbf{o}+t\mathbf{d}$ and compute the exact entry and exit distances for every intersected tetrahedron $t_k$, denoted as $t_k^{\mathrm{in}}<t_k^{\mathrm{out}}$, with $\mathbf{p}_k^{\mathrm{in}}=\mathbf{r}(t_k^{\mathrm{in}})$ and $\mathbf{p}_k^{\mathrm{out}}=\mathbf{r}(t_k^{\mathrm{out}})$. Along each ray, the contributions from these tetrahedra are accumulated in a front-to-back order using alpha compositing.

\textbf{SDF-based opacity.}
To render SDF within this framework, we map SDF values to an interval opacity as in NeuS~\cite{neus}. For each segment, we evaluate the piecewise-linear SDF at its endpoints
\begin{equation}
f_{\mathrm{prev}} = f(\mathbf{p}_k^{\mathrm{in}}),\qquad
f_{\mathrm{next}} = f(\mathbf{p}_k^{\mathrm{out}}).
\label{eq:sdf_endpoints}
\end{equation}
We map SDF values to an interval opacity using the logistic CDF $\Phi_s(x)=\big(1+e^{-s x}\big)^{-1}$, as in NeuS~\cite{neus}, where $s$ is the learned inverse standard deviation controlling surface sharpness.
\begin{equation}
\alpha_k
=
\max\!\left(\frac{\Phi_s(f_{\mathrm{prev}})-\Phi_s(f_{\mathrm{next}})}{\Phi_s(f_{\mathrm{prev}})},
0\right).
\label{eq:sdf_to_alpha}
\end{equation}

\textbf{Tetrahedron color and compositing.}
We then compute the tetrahedron color using the linear field from Sec.~\ref{sec:method_representation},
$\mathbf{c}_k(\mathbf{p},\mathbf{d})=\mathbf{c}_k^0(\mathbf{d})+\nabla \mathbf{c}_k\cdot(\mathbf{p}-\mathbf{O}_k)$.
We evaluate colors at the segment endpoints and use the segment-average color
\begin{equation}
\bar{\mathbf{c}}_k=\tfrac{1}{2}\Big(\mathbf{c}_k(\mathbf{p}_k^{\mathrm{in}},\mathbf{d})+\mathbf{c}_k(\mathbf{p}_k^{\mathrm{out}},\mathbf{d})\Big).
\label{eq:segment_color}
\end{equation}
The final pixel color is accumulated by standard alpha compositing in depth order
\begin{equation}
\mathbf{C}=\sum_k T_k\,\alpha_k\,\bar{\mathbf{c}}_k,\qquad
T_k=\prod_{l<k}\big(1-\alpha_l\big).
\label{eq:alpha_compositing}
\end{equation}

\textbf{Tetrahedron depth and normal.}
  Rendering other properties mirrors color compositing by replacing the color term with the target modality; we reuse the same
  weights $T_k\alpha_k$ for depth and normals. For depth, we assign each intersected tetrahedron a representative depth as the
  segment midpoint
  \begin{equation}
  z_k=\tfrac{1}{2}\big(t_k^{\mathrm{in}}+t_k^{\mathrm{out}}\big),
  \end{equation}
  and alpha‑composite it along the ray. For normals, we assume a constant normal within each tetrahedron. Since the SDF is linearly
  interpolated inside $t_k$, its gradient is constant; for
  $f(\mathbf{x})=\sum\limits_{j=1}^4 \lambda_j(\mathbf{x})\, f_{k,j}$,
  \begin{equation}
  \nabla f_k=\sum_{j=1}^4 f_{k,j}\,\nabla \lambda_j(\mathbf{x}),\qquad
  \mathbf{n}_k=\nabla f_k/\|\nabla f_k\|_2.
  \end{equation}
  The normal is obtained by alpha compositing and normalization.

\subsection{Surface-Centric Adaptive Strategy}
\label{sec:method_adaptive}

We use a surface-centric refinement strategy, together with culling and pruning for efficiency and compactness.

  \textbf{Densification.}
  To concentrate resolution near the target geometry, surface-crossing tetrahedra with mixed-sign vertex SDF values are ranked by
  circumradius, and the top-$k$ (5\% of all tetrahedra) are split by inserting a vertex at the tetrahedron centroid before re-tetrahedralizing.
  This allocates capacity to the evolving zero level set where detail is needed.

  We also adopt Radiance Meshes'~\cite{radiancemeshes} error-driven densification based on SSIM and total-variance scores aggregated per
  tetrahedron from multi-view residuals. Tetrahedra with the highest scores are split, and new
  vertices are inserted using the two most erroneous views by intersecting their mean rays, with a barycentric fallback
  when the intersection is degenerate.

\textbf{Culling.}
Culling is a non-destructive operation: tetrahedra are retained in the grid but omitted during rendering. A conservative opacity upper bound is obtained by pairing the largest and smallest vertex SDF values along a segment:
\begin{equation}
O_{\max}
\;=\;
1-\frac{\Phi_s(f_{\min})}{\Phi_s(f_{\max})}.
\end{equation}
Here $f_{\min}=\min_{i\in\{1,2,3,4\}} f_i$ and $f_{\max}=\max_{i\in\{1,2,3,4\}} f_i$, where $f_i$ are the SDF values at the tetrahedron's four vertices. A tetrahedron is skipped when $\min_{i\in t}|f_i|>\ell(s)$ and $O_{\max}<0.1$. The band width that contains 99\% of the logistic CDF mass is $\ell(s)\approx 2\ln 199/s$~\cite{svrecon}, so it contracts around the zero level set as the learnable parameter $s$ increases during optimization.

\textbf{Pruning.}
Pruning is destructive and keeps the representation compact. For each tetrahedron $t$, we record its peak contribution as the maximum alpha-compositing weight over pixels and views,
\begin{equation}
c_t = \max_{i,\pi} w_{t,i}^{\pi},
\end{equation}
and aggregate it to vertices by $c_v=\max_{t\ni v} c_t$, where $w_{t,i}^{\pi}=T_{t,i}^{\pi}\alpha_{t,i}^{\pi}$ is the per-pixel alpha-compositing weight in Sec.~\ref{sec:method_rendering}. A vertex $v$ is pruned when it is both low-contribution and far from the surface band,
\begin{equation}
c_v < \tau_c \quad \text{and} \quad |f(v)| > \ell(s),
\end{equation}
where $f(v)$ is the SDF at vertex $v$, $\tau_c$ is a contribution threshold. This is applied only to internal vertices. We then remove associated vertices/tetrahedra and re-tetrahedralize.

\subsection{Optimization}
      \label{sec:method_optimization}

      We optimize the parameters of the shared hash-encoded SDF and appearance predictors together with the tetrahedral vertices, under a combined photometric and geometric objective that keeps the extracted mesh
      consistent with the SDF. Each iteration renders the tetrahedral field, extracts a mesh with differentiable Marching Tetrahedra, and rasterizes the mesh to obtain depth/normal maps for geometric supervision. The total loss is
      \begin{equation}
      \mathcal{L}
      \;=\;
      \mathcal{L}_{\text{rgb}}
      \;+\;
      \lambda_{\text{mesh}}\mathcal{L}_{\text{mesh}}
      \;+\;
      \lambda_{\text{field}}\mathcal{L}_{\text{field}}.
      \label{eq:total_loss}
      \end{equation}
      For appearance, we use an $\ell_1$ reconstruction term with SSIM on rendered colors $\mathbf{C}$ and ground truth $\mathbf{C}^*$,
      \begin{equation}
      \mathcal{L}_{\text{rgb}}
      =
      \|\mathbf{C}-\mathbf{C}^*\|_1
      \;+\;
      \lambda_{\text{ssim}}\bigl(1-\mathrm{SSIM}\bigr).
      \label{eq:rgb_loss}
      \end{equation}
       To link surface losses to the SDF, we extract a triangle mesh $\mathcal{M}$ as the zero level set using differentiable Marching
        Tetrahedra~\cite{marchingtetrahedra}. For each tetrahedron whose vertex SDFs have opposite signs, the surface
        intersects its edges; the resulting 3 (or 4) intersection points form 1 (or 2) triangles. We denote each intersection
        point as $\mathbf{v}_n$. For an edge between vertices $\mathbf{v}_i$ and $\mathbf{v}_j$ with opposite-sign SDF values
        $f_i$ and $f_j$, the intersection point is
        \begin{equation}
        \mathbf{v}_{n}
        =
        \frac{f_i\,\mathbf{v}_j-f_j\,\mathbf{v}_i}{f_i-f_j},
        \label{eq:mt_vertex}
        \end{equation}
        which is differentiable with respect to both SDF values and vertex positions, allowing gradients to flow from mesh
        losses to the SDF.
      We then enforce field and surface consistency by rendering depth and normals from both (i) the tetrahedral SDF and (ii) the mesh $\mathcal{M}$. Let $D(i)$ and $\tilde{\mathbf{N}}(i)$ denote the depth and normal rendered from the field, and $D_{\text{M}}(i)$ and $\mathbf{N}_{\text{M}}(i)$ those rendered from the mesh. The weights $\lambda_{\text{mesh}}$, $\lambda_{\text{field}}$, $\lambda_{\text{MD}}$, and $\lambda_{\text{MN}}$ balance photometric, field, and mesh terms. The mesh term is
      \begin{equation}
      \mathcal{L}_{\text{mesh}}
      =
      \lambda_{\text{MD}}\mathcal{L}_{\text{MD}}
      \;+\;
      \lambda_{\text{MN}}\mathcal{L}_{\text{MN}},
      \label{eq:mesh_consistency}
      \end{equation}
      with
      \begin{equation}
      \mathcal{L}_{\text{MD}}
      =
      \sum_i
      \log\!\left(1 + \left|D(i) - D_{\text{M}}(i)\right|\right)
      \label{eq:mesh_depth_consistency}
      \end{equation}
      \begin{equation}
      \mathcal{L}_{\text{MN}}
      =
      \sum_i \left(1 - \tilde{\mathbf{N}}(i) \cdot \mathbf{N}_{\text{M}}(i)\right)
      \label{eq:mesh_normal_consistency}
      \end{equation}
  We add a normal and depth consistency loss on the field rendering, where $\mathbf{N}_{D}(i)$ is estimated from the rendered depth map $D$ via finite differences, following~\cite{2dgs}:
  \begin{equation}
  \mathcal{L}_{\text{ND}}
  =
  \sum_i \left(1 - \tilde{\mathbf{N}}(i) \cdot \mathbf{N}_{D}(i)\right).
  \label{eq:normal_depth_consistency}
  \end{equation}
  Finally, we regularize the SDF with an Eikonal loss over all tetrahedral vertices. We use numerical gradients because hash-encoded fields are only piecewise smooth and higher-order autodiff can be noisy or unstable; finite differences provide a robust estimate for Eikonal regularization~\cite{igrls,neuralangelo}. At each vertex $\mathbf{x}$, we query the SDF and estimate its gradient via finite differences, e.g.,
  \begin{equation}
    \nabla_x f(\mathbf{x}) =
    \frac{f\!\left(\mathbf{x}+\boldsymbol{\epsilon}_x\right)-f\!\left(\mathbf{x}-\boldsymbol{\epsilon}_x\right)}{2\epsilon},
    \label{eq:eikonal_finite_diff}
    \end{equation}
    where $\boldsymbol{\epsilon}_x=[\epsilon,0,0]$, and $\nabla_y f(\mathbf{x})$ and $\nabla_z f(\mathbf{x})$ are computed
    similarly. We set $\epsilon$ to the cell size of the finest currently active hash-grid resolution.
      \begin{equation}
        \mathcal{L}_{\text{eik}}
        =
        \frac{1}{N} \sum_{i=1}^N (\| \nabla f(\mathbf{x}_i) \|_2 - 1)^2 ,
        \label{eq:eikonal}
    \end{equation}

    \begin{equation}
        \mathcal{L}_{\text{curv}}
        =
        \frac{1}{N} \sum_{i=1}^N \left| \nabla^2 f(\mathbf{x}_i) \right| \,.
        \label{eq:curv}
    \end{equation}
        Here $N$ is the number of tetrahedral vertices.
        The field term is
        \begin{equation}
        \mathcal{L}_{\text{field}}
        =
        \lambda_{\text{ND}}\mathcal{L}_{\text{ND}}
        \;+\;
        \lambda_{\text{eik}}\mathcal{L}_{\text{eik}}
        \;+\;
        \lambda_{\text{curv}}\mathcal{L}_{\text{curv}}.
        \label{eq:field_loss}
        \end{equation}
 
\section{Experiments}
\label{sec:experiments}
\begin{figure*}[htbp]
    \centering
    \includegraphics[width=1.0\linewidth]{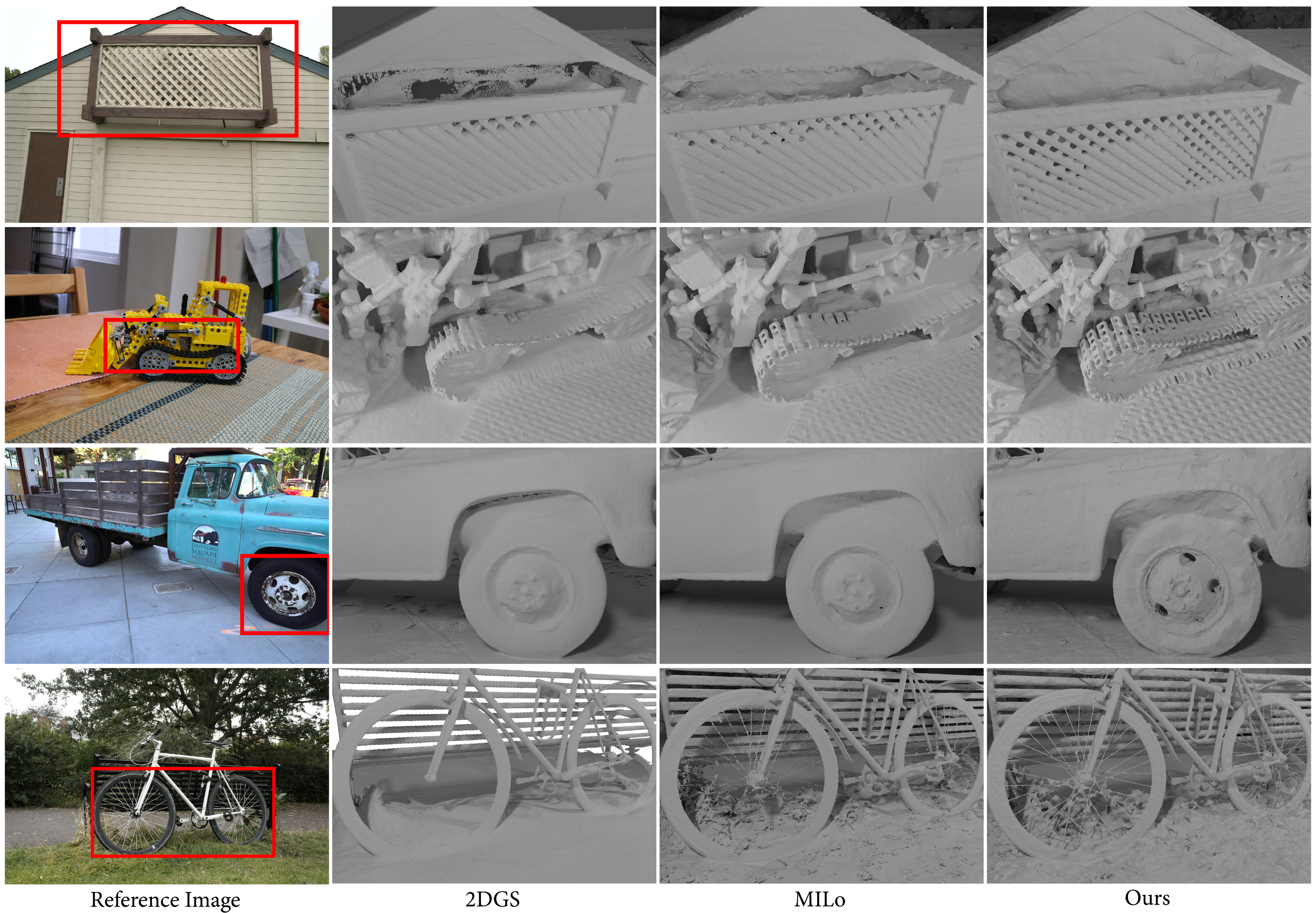}
    \caption{\textbf{Surface Reconstruction on the Tanks and Temples Dataset~\cite{tanksandtemples}} and Mip-NeRF 360 Dataset~\cite{mipnerf360}. Qualitative comparison on four scenes (Barn, Truck, Bicycle, Kitchen) with MILo~\cite{milo}, 2DGS~\cite{2dgs}, and our method. Our method converges to more accurate geometry while keeping the extracted meshes compact.
    }
    \label{fig:tnt}
\end{figure*}

\begin{figure*}[htbp]
    \centering
    \includegraphics[width=1.0\linewidth]{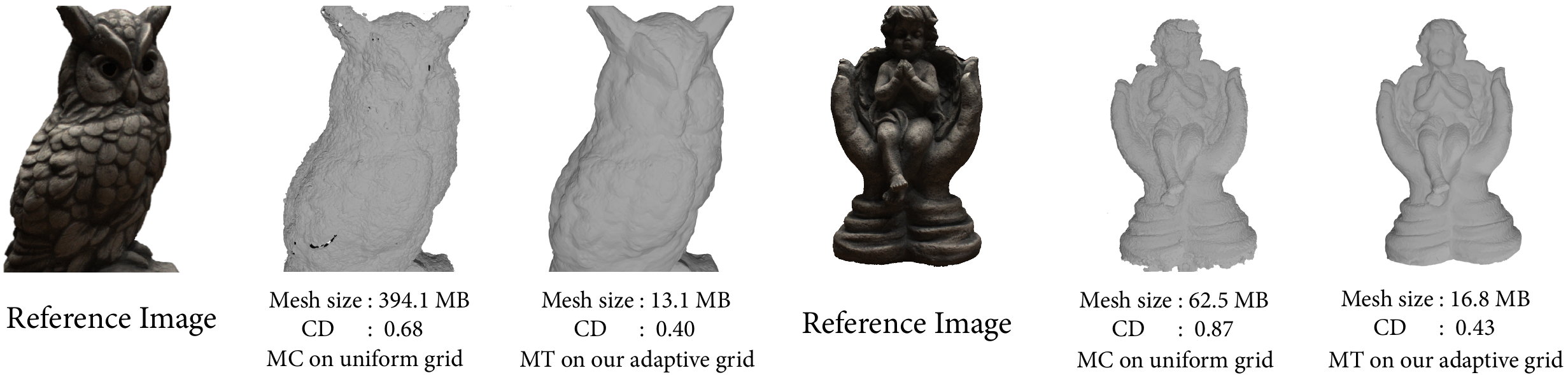}
    \caption{Comparison between Marching Tetrahedra (MT) ~\cite{marchingtetrahedra} on the adaptive tetrahedral grid and Marching Cubes (MC)~\cite{marchingcube} on a fixed-resolution grid using the same SDF optimized by SDFRaster. Despite the higher uniform resolution, Marching Cubes produces noisier surfaces, holes, and floaters. Our MT extraction yields lower Chamfer distance and a more compact mesh.
    }
    \label{fig:compare2}
\end{figure*}
\begin{figure*}[htbp]
    \centering
    \includegraphics[width=1.0\linewidth]{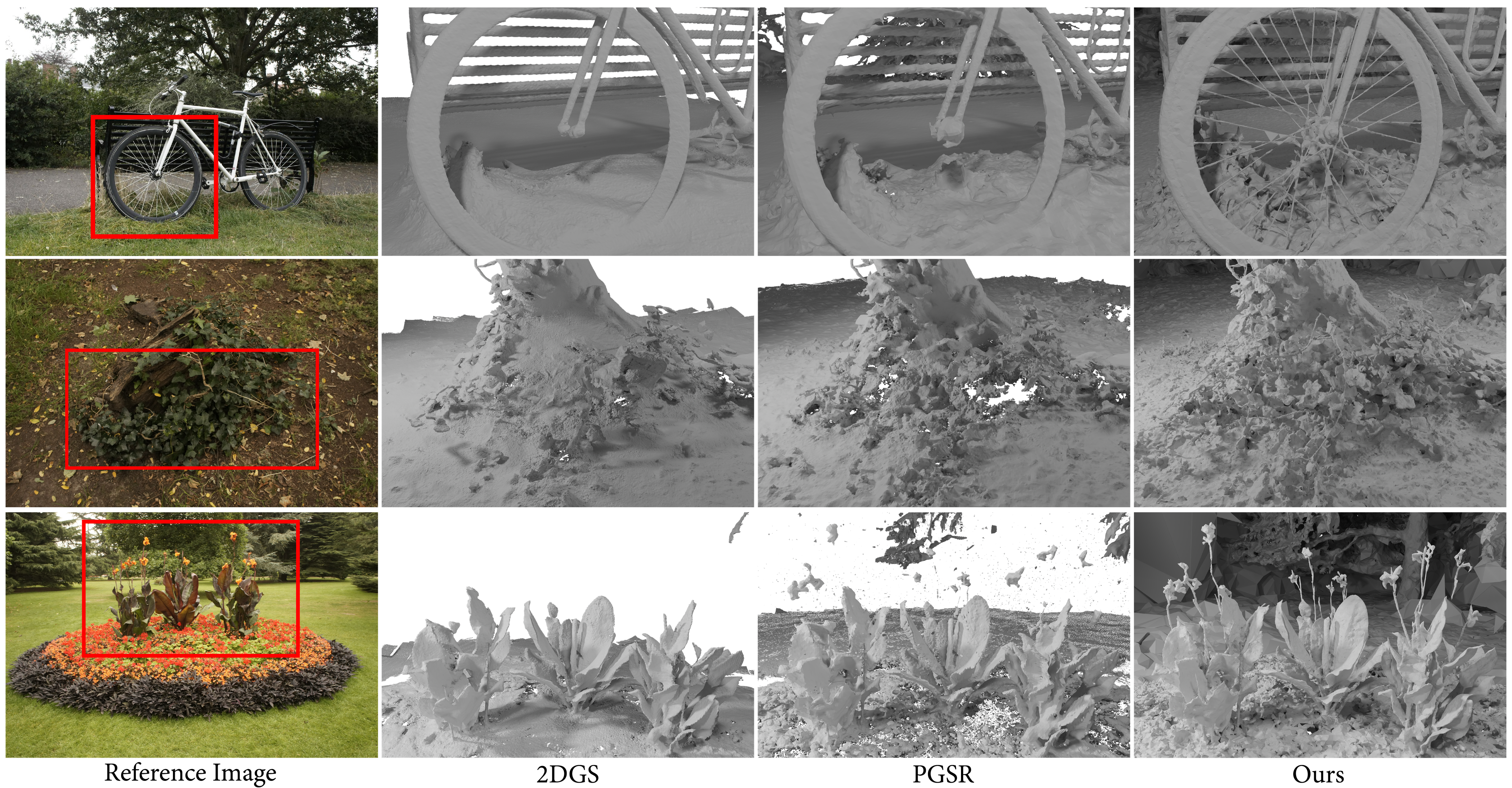}
    \caption{Comparison between 2DGS~\cite{2dgs} and PGSR~\cite{pgsr}, which use depth fusion for mesh extraction, and our SDF-based mesh extraction. 2DGS and PGSR render multi-view depth maps and fuse them with TSDF~\cite{tsdf}, a pipeline that is sensitive to view-dependent depth noise and often struggles with thin structures. In contrast, we extract the zero level set from the learned SDF on an adaptive tetrahedral grid, yielding more complete surfaces and better preserving fine details.
    }
    \label{fig:compare}
\end{figure*}
\setlength\tabcolsep{0.5em}
\begin{table*}[!ht]
\centering
\small
\caption{\textbf{Surface reconstruction metrics on the DTU dataset}. We report the Chamfer Distance across 15 scenes (lower is better). Implicit methods are listed above and explicit methods below. Best results for implicit methods are highlighted in blue; best results for explicit methods are highlighted in red.
}
\vspace{-0.2cm}
\resizebox{.98\textwidth}{!}{
\begin{tabular}{@{}llc*{15}{c}ccc@{}}
\toprule
\multicolumn{3}{c}{} & 24 & 37 & 40 & 55 & 63 & 65 & 69 & 83 & 97 & 105 & 106 & 110 & 114 & 118 & 122 & & Mean & Time \\
\cmidrule(lr){4-18}\cmidrule(lr){20-21}

\multirow{3}{*}{\rotatebox[origin=c]{90}{implicit}}
& VolSDF~\cite{volsdf} & & \tbestbis 1.14 & \sbestbis 1.26 & \sbestbis 0.81 & \tbestbis 0.49 & \tbestbis 1.25 & \tbestbis 0.70 & \tbestbis 0.72 & \sbestbis 1.29 & \tbestbis 1.18 & \bestbis 0.70 & \tbestbis 0.66 & \sbestbis 1.08 & \tbestbis 0.42 & \tbestbis 0.61 & \tbestbis 0.55 & & \tbestbis 0.86 & > 12h \\
& NeuS~\cite{neus} & & \sbestbis 1.00 & \tbestbis 1.37 & \tbestbis 0.93 & \sbestbis 0.43 & \sbestbis 1.10 & \sbestbis 0.65 & \sbestbis 0.57 & \tbestbis 1.48 & \sbestbis 1.09 & \tbestbis 0.83 & \sbestbis 0.52 & \tbestbis 1.20 & \sbestbis 0.35 & \sbestbis 0.49 & \sbestbis 0.54 & & \sbestbis 0.84 & > 12h \\
& Neuralangelo~\cite{neuralangelo} & & \bestbis 0.37 & \bestbis 0.72 & \bestbis 0.35 & \bestbis 0.35 & \bestbis 0.87 & \bestbis 0.54 & \bestbis 0.53 & \bestbis 1.29 & \bestbis 0.97 & \sbestbis 0.73 & \bestbis 0.47 & \bestbis 0.74 & \bestbis 0.32 & \bestbis 0.41 & \bestbis 0.43 & & \bestbis 0.61 & > 12h \\

\midrule

\multirow{7}{*}{\rotatebox[origin=c]{90}{explicit}}
& SuGaR~\cite{sugar} & & 1.47 & 1.33 & 1.13 & 0.61 & 2.25 & 1.71 & 1.15 & 1.63 & 1.62 & 1.07 & 0.79 & 2.45 & 0.98 & 0.88 & 0.79 & & 1.32 & 52m \\
& SVRaster~\cite{svraster} & & 0.88 & 2.12 & 0.78 & 0.53 & \sbest 0.87 & \tbest 1.01 & 2.10 & \best 1.08 & \best 1.20 & 1.45 & 1.08 & \best 0.85 & \tbest 0.43 & \sbest 0.60 & 0.93 & & 1.06 & \phantom{0}5m \\
& 2DGS~\cite{2dgs} & & 0.71 & 0.91 & 0.77 & \tbest 0.43 & 1.49 & 1.08 & 0.93 & 1.45 & 1.42 & 0.95 & 0.84 & 1.59 & 0.79 & 0.83 & 0.57 & & 0.98 & \phantom{0}9m \\
& GOF~\cite{gof} & & \tbest 0.69 & \tbest 0.89 & \tbest 0.48 & 0.52 & 1.67 & 1.03 & \sbest 0.82 & \tbest 1.29 & 1.42 & \tbest 0.82 & \sbest 0.61 & 1.28 & 0.55 & \tbest 0.62 & \tbest 0.48 & & \tbest 0.88 & 55m \\
& MILo~\cite{milo} & & \best 0.43 & \sbest 0.74 & \best 0.32 & \sbest 0.36 & \best 0.80 & \sbest 0.75 & \best 0.69 & 1.34 & \sbest 1.29 & \sbest 0.72 & \tbest 0.67 & \sbest 0.93 & \best 0.34 & 0.74 & \sbest 0.47 & & \sbest 0.71 & 61m \\
& Radiance Meshes~\cite{radiancemeshes} & & 3.17 & 5.02 & 1.31 & 0.97 & 4.93 & 4.71 & 2.32 & 4.73 & 2.68 & 2.80 & 2.48 & 2.38 & 1.85 & 2.60 & 1.67 & & 2.91 & 79m \\
& \textbf{Ours} & & \sbest 0.43 & \best 0.67 & \sbest 0.44 & \best 0.33 & \tbest 0.93 & \best 0.68 & \tbest 0.89 & \sbest 1.17 & \tbest 1.30 & \best 0.64 & \best 0.53 & \tbest 1.05 & \sbest 0.37 & \best 0.43 & \best 0.40 & & \best 0.68 & 98m \\

\bottomrule
\end{tabular}
}
\label{tab:surface_metrics_dtu}
\vspace{-0.2cm}
\end{table*}

\begin{figure*}[htbp]
    \centering
    \includegraphics[width=1.0\linewidth]{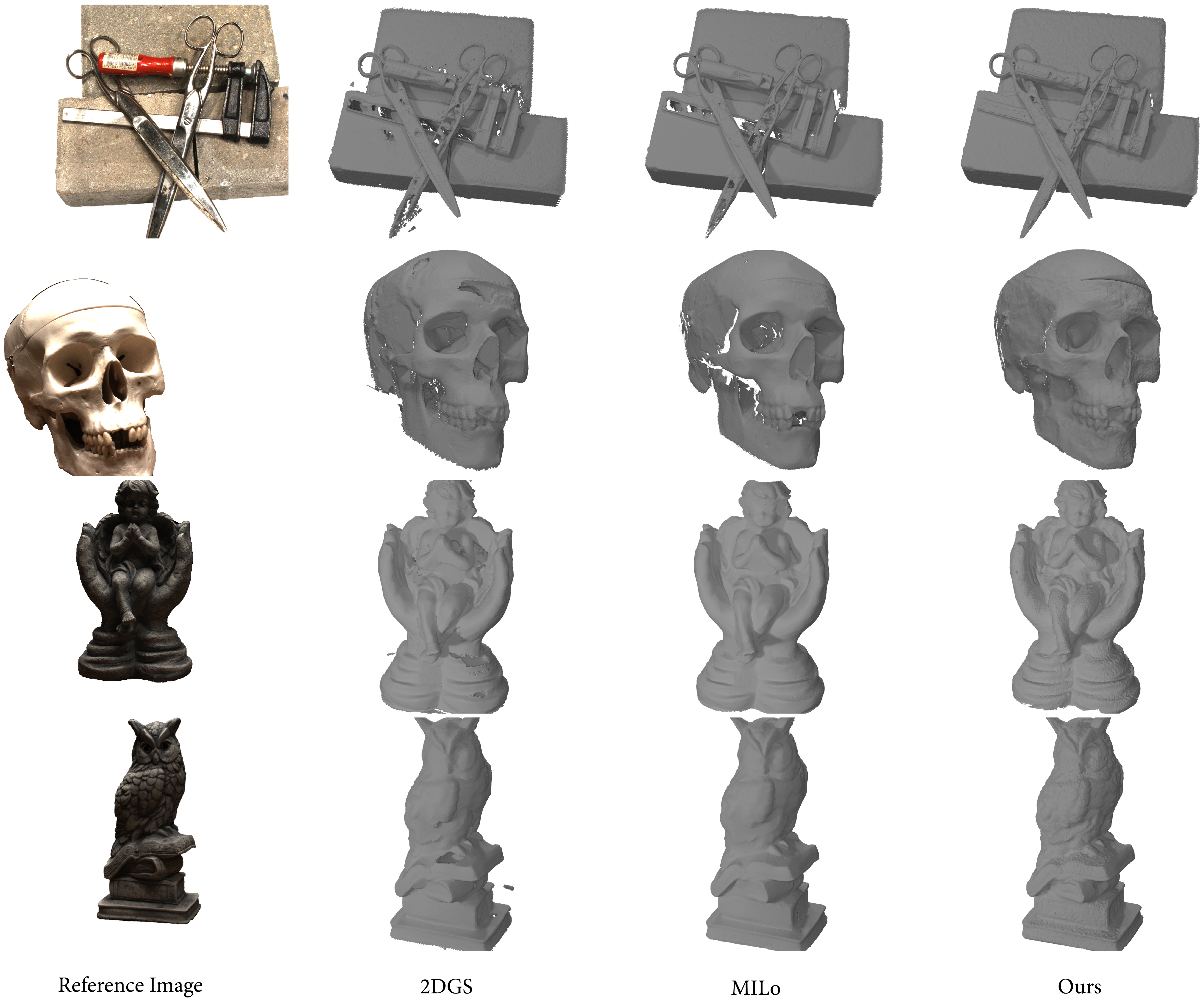}
    \caption{\textbf{Surface Reconstruction on the DTU Dataset~\cite{dtu}}. Qualitative comparison on four scans (Scan 37, Scan 65, Scan 118, Scan 122) with MILo~\cite{milo}, 2DGS~\cite{2dgs}, and our method.
    }
    \label{fig:dtu}
\end{figure*}

\begin{table}[t]
\centering
\small
\setlength{\tabcolsep}{3.5pt}
\caption{\textbf{Quantitative comparison on the Tanks and Temples dataset}. We report the F1-score and output mesh size (MB) across six scenes (higher is better). All results are evaluated with the official evaluation scripts. Best results for implicit methods are highlighted in blue; best results for explicit methods are highlighted in red. Abbreviations: Cat. (Caterpillar), Court. (Courthouse), Ign. (Ignatius), Meet. (Meetingroom), and RadMesh (Radiance Meshes~\cite{radiancemeshes}).}
\vspace{-0.15cm}
\resizebox{0.95\linewidth}{!}{
\begin{tabular}{@{}lcccccccc@{}}
\toprule
Method & Barn & Cat. & Court. & Ign. & Meet. & Truck & Mean & \multicolumn{1}{c}{Size} \\
\midrule
NeuS         & \tbestbis 0.29 & \sbestbis 0.29 & \sbestbis 0.17 & \sbestbis 0.83 & \sbestbis 0.24 & \sbestbis 0.45 & \sbestbis 0.38 & - \\
GeoNeuS      & \sbestbis 0.33 & \tbestbis 0.26 & \tbestbis 0.12 & \tbestbis 0.72 & \tbestbis 0.20 & \tbestbis 0.45 & \tbestbis 0.35 & - \\
Neuralangelo & \bestbis 0.70  & \bestbis 0.36  & \bestbis 0.28  & \bestbis 0.89  & \bestbis 0.32  & \bestbis 0.48  & \bestbis 0.50  & - \\
\midrule
SuGaR        & 0.14            & 0.16            & 0.08            & 0.33            & 0.15            & 0.26            & 0.19            & - \\
2DGS         & 0.38            & 0.18            & 0.13            & 0.30            & 0.16            & 0.30            & 0.21            & 658.2 \\
GOF          & \sbest 0.55     & \sbest 0.38     & \sbest 0.29     & \tbest 0.73     & \best 0.26      & \sbest 0.56     & \sbest 0.46     & 605.4 \\
MILo         & \best 0.55      & \best 0.38      & \tbest 0.26     & \best 0.74      & \sbest 0.25     & \best 0.60      & \best 0.46      & 176.3 \\
RadMesh & 0.30 & 0.23 & 0.14 & 0.35 & 0.04 & 0.28 & 0.22 & \phantom{0}42.3 \\
\textbf{Ours} & \tbest 0.54    & \tbest 0.33     & \best 0.30      & \sbest 0.73     & \tbest0.22            & \tbest0.48            & \tbest 0.43     & 186.2 \\
\bottomrule
\end{tabular}
}
\label{tab:surface_metrics_tandt}
\vspace{-0.2cm}
\end{table}
\begin{table}[t]
    \centering
    \small
    \setlength{\tabcolsep}{2.5pt}
    \caption{\textbf{Quantitative results for novel view synthesis on the Mip-NeRF360 dataset}. We report PSNR, SSIM, and LPIPS for outdoor and indoor scenes separately. Abbreviations: RadMesh (Radiance Meshes~\cite{radiancemeshes})}
    \vspace{-0.2cm}
    \resizebox{0.8\linewidth}{!}{
    \begin{tabular}{@{}lcccccc@{}}
    \toprule
     & \multicolumn{3}{c}{Outdoor Scenes} & \multicolumn{3}{c}{Indoor Scenes} \\
    \cmidrule(lr){2-4} \cmidrule(lr){5-7}
     Method & PSNR $\uparrow$ & SSIM $\uparrow$ & LPIPS $\downarrow$
     & PSNR $\uparrow$ & SSIM $\uparrow$ & LPIPS $\downarrow$ \\
    \midrule
    2DGS
    & 24.15 & 0.702 & \sbest 0.288
    & \sbest 30.02 & 0.909 & \sbest 0.213 \\
    MILo
    & \sbest 24.23 & \best 0.740 & \best 0.264
    & 28.93 & \sbest 0.919 & \best 0.185 \\
    RadMesh
    & \best 24.38 & \sbest 0.721 & \tbest 0.292
    & \best 30.61 & \best 0.920 & 0.252 \\
    \textbf{Ours}
    & \tbest 24.22 & \tbest 0.715 & 0.293
    & \tbest 29.58 & \tbest 0.918 & \tbest 0.233 \\
    \bottomrule
    \end{tabular}
    }
    \label{tab:nvs_metrics_mip360}
    \vspace{-0.3cm}
\end{table}

\subsection{Implementation Details}

Our implementation is built in PyTorch with SlangD shaders on top of the Radiance Meshes~\cite{radiancemeshes} codebase.
We follow the training schedule in Sec.~\ref{sec:method}, use the adaptive densification and pruning strategy in Sec.~\ref{sec:method_adaptive}, and extract meshes with differentiable Marching Tetrahedra~\cite{marchingtetrahedra} during training.
 Unless otherwise specified, we set the loss weights to $\lambda_{\text{eik}}=0.01$, $\lambda_{\text{curv}}=5\times10^{-6}$, $\lambda_{\text{ND}}=\lambda_{\text{MD}}=\lambda_{\text{MN}}=0.05$, and $\lambda_{\text{ssim}}=0.2$.
 We train on each dataset for 18,000 iterations. DTU~\cite{dtu} experiments are run on a single NVIDIA RTX 4090, while TnT~\cite{tanksandtemples} and Mip-NeRF 360~\cite{mipnerf360} are run on a single NVIDIA L40. Average training time is about 98 minutes per DTU scene and 6 hours per TnT scene.
We densify every 500 iterations from 2,000 to 16,000 and prune every 500 iterations from 4,000 to 15,000 using the criteria in Sec.~\ref{sec:method_adaptive}.
For better surface reconstruction on TnT, we increase the mesh loss weight by $5\times$. On complex unbounded scenes such as TnT, we disable the Eikonal regularizer to improve convergence; an ablation is provided in Sec.~\ref{sec:ablation}. 

\subsection{Comparison}

 We evaluate on DTU~\cite{dtu} and Tanks and Temples (TnT)~\cite{tanksandtemples} for surface reconstruction and include qualitative mesh comparisons on Mip-NeRF 360~\cite{mipnerf360}; we use the standard 15 DTU scenes and the common six-scene TnT subset (Barn, Caterpillar, Ignatius, Courthouse, Meetingroom, Truck). Mip-NeRF 360 provides unbounded scenes without ground-truth geometry, so we present qualitative mesh comparisons there.

Following standard practice, we report Chamfer distance (CD) on DTU and F1-score on TnT; for TnT, F1-score combines precision and recall at the standard distance threshold, and for DTU we follow the official evaluation script and report the average of accuracy and completeness as Chamfer distance. We cull extracted meshes with camera masks on DTU and visibility filtering on TnT before evaluation. We compare our method with explicit and implicit approaches for surface reconstruction from multi-view images. We follow each paper's mesh extraction procedure and use official implementations and recommended hyperparameters when reproducing results.

Tab.~\ref{tab:surface_metrics_dtu} and Tab.~\ref{tab:surface_metrics_tandt} summarize mesh reconstruction quality on DTU and TnT. On DTU, SDFRaster improves Chamfer distance over splatting-based baselines that rely on post-processing methods to extract meshes, and Fig.~\ref{fig:dtu} shows fewer holes and cleaner surfaces across four scenes. On TnT, SDFRaster achieves higher F1-scores while keeping meshes compact, indicating that the adaptive tetrahedral grid allocates resolution where it is most needed; Fig.~\ref{fig:tnt} shows better thin-structure recovery and finer detail on two TnT scenes and two Mip-NeRF 360 scenes. Compared with implicit SDF baselines, SDFRaster achieves competitive accuracy while retaining the efficiency and scalability of explicit rendering. Tab.~\ref{tab:nvs_metrics_mip360} further reports the novel view synthesis results on Mip-NeRF 360. Although surface reconstruction is our main goal, our method still maintains competitive rendering quality in terms of PSNR, SSIM, and LPIPS compared with previous approaches.

We also compare against 2DGS and PGSR, which render per-view depth maps and extract meshes via TSDF fusion; this depth fusion pipeline is sensitive to multi-view depth consistency and struggles with thin structures, often producing holes, missing regions, or overly dense meshes. Fig.~\ref{fig:compare} and Tab.~\ref{tab:surface_metrics_tandt} show that SDFRaster yields cleaner topology and preserves sharp edges and thin structures while using fewer mesh vertices and faces. TSDF fusion relies on a uniform voxel grid whose resolution and memory trade-off limits scalability in large scenes, while our adaptive tetrahedral grid allocates capacity where needed and enables background reconstruction without sacrificing foreground detail. By learning a globally consistent SDF and extracting its zero level set directly, we avoid depth-fusion artifacts caused by view-dependent depth noise and obtain more complete surfaces.

\begin{table}[t]
    \centering
    \small
    \setlength{\tabcolsep}{4pt}
    \renewcommand{\arraystretch}{1.05}
    \caption{\textbf{Ablation studies on TnT and DTU.} Left: average F1-score on Tanks and Temples. Right: average Chamfer Distance on DTU.}
    \label{tab:loss_ablation}
    \vspace{-0.15cm}
    \resizebox{0.95\linewidth}{!}{
    \begin{tabular}{@{}c@{\hspace{1.2em}}c@{}}
    \begin{tabular}[t]{@{}lc@{}}
        \multicolumn{2}{c}{\textbf{\strut Tanks and Temples}}\\[2pt]
        \toprule
        Method & F1-score $\uparrow$ \\
        \midrule
        Baseline & 0.3358 \\
        Baseline + $\mathcal{L}_{\text{MD}}$ & 0.3519 \\
        Baseline + $\mathcal{L}_{\text{ND}}$ & 0.4137 \\
        Baseline + $\mathcal{L}_{\text{mesh}}$ & 0.3908 \\
        Full model (w/ all losses) & 0.4023 \\
        \midrule
        Default setting (w/o $\mathcal{L}_{\text{eik}}$) & \textbf{0.4289} \\
        \bottomrule
    \end{tabular}
    &
    \begin{tabular}[t]{@{}lc@{}}
        \multicolumn{2}{c}{\textbf{\strut DTU}}\\[2pt]
        \toprule
        Method & CD $\downarrow$ \\
        \midrule
        w/o $\mathcal{L}_{\text{curv}}$ & 0.69 \\
        w/o $\mathcal{L}_{\text{eik}}$ & 0.72 \\
        w/o densification & 0.79 \\
        w/o culling & 0.68 \\
        w/o pruning & 0.71 \\
        \midrule
        \textbf{Ours} (w/ all) & \textbf{0.68} \\
        \bottomrule
    \end{tabular}
    \end{tabular}
    }
    \vspace{-0.4cm}
\end{table}

\subsection{Ablation Study}
\label{sec:ablation}

We conduct ablations on both TnT and DTU to evaluate the contributions of our geometry consistency losses, the Eikonal regularizer, and the adaptive refinement strategy. Tab.~\ref{tab:loss_ablation} summarizes the quantitative impact on TnT. Adding $\mathcal{L}_{\text{MD}}$ yields a modest F1 improvement, while $\mathcal{L}_{\text{ND}}$ brings a larger gain, indicating that normal and depth consistency more effectively stabilizes the field geometry. The mesh consistency term also improves F1, and the combination of these losses further improves performance, showing the mesh and field coupling and field-level normal constraints are complementary.

On bounded object-centric scenes, the Eikonal term improves reconstruction quality on DTU (Tab.~\ref{tab:loss_ablation}). The adaptive refinement strategy is also important: densification and pruning improve reconstruction quality, while culling mainly benefits rendering efficiency. On complex unbounded scenes such as TnT, the Eikonal regularizer substantially hinders convergence (Tab.~\ref{tab:loss_ablation}). We therefore disable $\mathcal{L}_{\text{eik}}$ on TnT. Although this removes the strict signed-distance constraint on the learned field, the zero level set remains stable for surface reconstruction under image supervision.

Since our SDF can be queried at arbitrary locations, we can also extract surfaces on a uniform grid with Marching Cubes. We compare this to Marching Tetrahedra on our adaptive tetrahedral grid in Fig.~\ref{fig:compare2}. Despite the higher uniform resolution, Marching Cubes yields noisier surfaces, local holes and floaters, while our extraction preserves fine details with a more compact mesh, indicating that the SDF optimization is tightly coupled to the tetrahedral grid.
\section{Conclusion}
We propose SDFRaster, a rasterizable signed distance field framework for end-to-end mesh reconstruction from multi-view images. The method is built on a rasterizable, multi-view consistent geometry representation that enables rapid optimization and direct mesh extraction during optimization, coupling distance field optimization with surface fidelity via differentiable Marching Tetrahedra. It further incorporates a surface-centric adaptive strategy that concentrates capacity near the zero-level set to produce compact meshes.

Experiments show that SDFRaster can produce accurate surfaces and compact meshes, and extends to unbounded scenes. We believe that rasterizable signed distance fields are a promising direction for scalable, fast, and high-fidelity surface reconstruction, providing a principled alternative to volumetric-primitive methods that lack multi-view consistent surface geometry and to implicit methods that require costly dense sampling during optimization.

One limitation of SDFRaster is that it uses a hash-encoded MLP to parameterize the SDF, rather than explicitly storing the SDF value for each vertex. While this network parameterization provides the smoothness inductive bias inherent in MLPs, it also introduces a nontrivial evaluation overhead compared to a purely explicit splatting pipeline. Future directions worth exploring include explicitly storing the SDF on a tetrahedral mesh and replacing the implicit MLP prior with direct smoothness regularization of the field, which would eliminate the need for network queries and potentially significantly accelerate training.

\begin{acks}
This research was supported by the National Natural Science Foundation of China (No.62272433), Anhui Provincial Natural Science Foundation (No.2508085ZD011), Key Science and Technology Project of Anhui Province (No.202523o09050004) and the Fundamental Research Funds for the Central Universities.
\end{acks}

% \input{table_compare}

\bibliographystyle{ACM-Reference-Format}
\bibliography{main,jk}


\end{document}